\documentclass[traditabstract]{aa}
\usepackage{graphicx}
\usepackage{natbib}
\usepackage{txfonts}
\usepackage{epsfig}
\usepackage{times}

\begin{document}

\title{Extragalactic gamma-ray background from AGN winds and star-forming galaxies in cosmological galaxy formation models}

\authorrunning{A. Lamastra, et al.}

  \author{A. Lamastra\inst{1,2}, N. Menci\inst{1}, F. Fiore\inst{1}, L. A. Antonelli\inst{1,2}, S. Colafrancesco\inst{3}, D. Guetta\inst{1,4}, A. Stamerra\inst{5,6}}
 \offprints{alessandra.lamastra@oa-roma.inaf.it}
   \institute{$^1$ INAF - Osservatorio Astronomico di Roma, via di Frascati 33, 00078 Monte Porzio Catone, Italy\\
   	$^2$ SSDC--ASI, Via del Politecnico, 00133 Roma, Italy\\   	
   	$^3$ School of Physics, University of the Witwatersrand, Private Bag 3, 2050-Johannesburg, South Africa\\
   	$^4$ Department of Physics and Optical Engineering, ORT Braude College, Karmiel 21982, Israel\\
   	$^5$ INAF - Osservatorio Astrofisico di Torino, via Osservatorio, 20, 10025 Pino Torinese, Italy \\
   	 $^6$  Scuola Normale Superiore, Piazza dei Cavalieri 7, 56126 Pisa, Italy \\}  
   \date{Received ; Accepted }
   \abstract{
  We derive the contribution to the extragalactic gamma-ray background (EGB) from  AGN winds and star-forming galaxies by including a physical model for  the  $\gamma$-ray emission produced by  relativistic protons  accelerated by AGN-driven and supernova-driven shocks into  a  state-of-the-art semi-analytic model  of galaxy formation. This is based on galaxy interactions as triggers of AGN accretion and starburst activity and on expanding  blast wave  as the mechanism to communicate outwards the energy injected into the interstellar medium by the active nucleus. We compare the model predictions with the latest measurement of the  EGB spectrum performed by the {\it Fermi}-LAT in the range between 100 MeV and 820 GeV. We find that AGN winds can provide $ \sim $35$\pm$15\% of the observed EGB in the energy interval $E_{\gamma}=$0.1-1 GeV,  for $ \sim $73$\pm$15\% at $E_{\gamma}=$1-10 GeV, and for $ \sim $60$\pm$20\% at   $E_{\gamma}\gtrsim$10 GeV.
The AGN wind contribution to the EGB is predicted to be larger by a factor of $ \sim $3-5 than that provided by  star-forming galaxies (quiescent plus starburst) in the hierarchical clustering scenario.
The cumulative $\gamma-$ray emission from AGN winds and blazars can account for the amplitude and spectral shape of the EGB, assuming the standard acceleration theory, and AGN wind parameters  that agree  with observations.
  We also compare the model prediction for the cumulative neutrino background  from AGN winds with the most recent IceCube data. We find that for AGN winds with  accelerated proton spectral index $p$=2.2-2.3, and taking into account internal absorption of $\gamma$-rays,  the {\it Fermi}-LAT  and IceCube data could be  reproduced simultaneously.

 \keywords{galaxies: active -- galaxies:  evolution --   --galaxies:  formation 
 }}
 
\titlerunning{Contribution from AGN winds and starburst galaxies to the EGB }

 \maketitle
 
\section{Introduction}

The Extragalactic Gamma-ray background (EGB) represents a superposition  of all $\gamma$-ray sources, both individual and diffuse,  from  the Milky Way to the edge of the observable universe, and provides a  view of the high-energy  processes in the universe.
Here we consider the total  $\gamma$-ray photon flux  produced outside of the Milky Way, including both resolved and unresolved sources. Indeed, the diffuse Galactic emission  produced by the interaction of Galactic cosmic rays (CR), mainly protons and electrons, with the Galactic interstellar medium (ISM) and interstellar radiation field, is comparable to the EGB intensity and represents a strong foreground to the EGB measurement. The latter have been recently measured by  the Large Area Telescope (LAT,  \citealt{Atwood09} )  on board the {\it Fermi Gamma-ray Space Telescope} ({\it Fermi}), in the range between 100 Mev and 820 GeV \citep{Ackermann15}. The EGB spectrum  is well described by a power-law with exponential cut-off having a spectral index of $\sim$2.3 and cut-off energy greater than 300 GeV.

How much different source classes contribute to the EGB remains one of the main unanswered questions of $\gamma$-ray astrophysics.
Well-established astrophysical populations, whose brightest members have been robustly detected, represent guaranteed components to the EGB. Among these, the extragalactic components  are blazars, radio galaxies, and star-forming galaxies \citep[see][for a review]{Fornasa15}.

Blazars are among the brightest $\gamma$-ray emitters in the sky.  They account for $ \sim $50$^{+12}_{-11}$\% of the EGB in the energy interval $E_{\gamma}\lesssim$10 GeV,  and for $ \sim $85$^{+15}_{-21}$\% at $E_{\gamma}\gtrsim$10 GeV \citep{Ajello15}. They are interpreted as active galactic nuclei (AGN) with the relativistic jet directed towards the observer.  The $\gamma$-ray emission in blazars is produced by inverse Compton (IC) scattering of the electrons accelerated in the jet and either the synchrotron photons emitted by the same leptonic population (synchrotron-self Compton),   or from accretion disk photons (external Compton). 

According to the AGN unification model \citep{Antonucci85} the viewing angle discriminates among blazars and radio galaxies. With no doppler boost, radio galaxies are expected to be less bright but more abundant than blazars (blazars represent $\sim$10\% of the AGN population), making them potentially important contributors to the EGB. However their contribution to EGB is not well constrained ranging from $\sim$7\% to $\sim$30\% of the EGB intensity measured by {\it Fermi}-LAT at  $E_{\gamma}\lesssim$10 GeV   \citep{Inoue11,DiMauro14,Wang16_gamma}.

Recently, also the AGN population that do not exhibit relativistic jets, have been considered as possible astrophysical contributors to the EGB  \citep{Wang16_gamma}.  In fact, several observational evidence indicate that AGN produce wide-angle winds with velocities of $v\sim$0.1-0.3 $c$ (e.g. \citealt{Chartas02,Pounds03,Reeves03,Tombesi10,Tombesi15}). The shocks produced by the interaction of AGN winds with the ambient medium are expected to accelerate particles to relativistic energies. The interactions of shock-accelerated particles with surrounding ISM and interstellar radiation field can produce non-thermal emission in the  $\gamma$-ray band \citep{Nims15,Wang16_gamma,Lamastra16}.

The same emission mechanisms are expected to produce $\gamma$-rays in star-forming galaxies.
In this case, the shocks are produced by supernovae (SN) explosions following star formation. Two modes of star formation  have been observationally identified:  a quiescent mode  where the star formation is extended over the whole galactic disk and occurs on time scales of (1-2) Gyr; and  a starburst mode where the star formation is concentrated in the dense, nuclear region of  galaxies,  and it is sustained at an enhanced rate in comparison to quiescently star-forming galaxies. 
There are several studies that derive the contribution to the EGB from star-forming galaxies \citep[e.g.][]{Fields10,Stecker11,Makiya11,Ackermann12,Chakraborty13,Lacki14,Tamborra14}. The studies that analyze  quiescent and starburst galaxies  separately find that the starburst contribution is always comparatively minor and the total (quiescent plus starburst) $\gamma$-ray emission is  between 10\%-50\%  of the EGB intensity.

However, large uncertainties remain for the contribution to the EGB of the above source classes. In this paper we present improved modelling of the integrated $\gamma$-ray emission from AGN winds and star-forming galaxies by incorporating a physical model for  the  $\gamma$-ray emission produced by particles accelerated in AGN-driven and SN-driven shocks \citep{Lamastra16} into a  state-of-the-art semi-analytic model (SAM)  of galaxy formation  \citep{Menci14}.
Our SAM includes a physical description of starburst and AGN activities triggered  by galaxy interactions during their merging histories, and is ideally suited for this goal as it has been tested against  several observational properties  of the AGN and galaxy population both in the local and high redshift universe, and in different electromagnetic bands (e.g. \citealt{Menci05, Menci06,Lamastra10,Lamastra13a,Lamastra13b,Menci14,Gatti15}). Moreover, galaxy and AGN number densities, and galaxy properties that determine the $\gamma$-ray emission, like gas mass, star formation rate (SFR), and AGN bolometric luminosity, can be calculated self-consistently by our SAM. This represent an advantage of the semi-analytic  approach  with respect to previous studies  based on  parametric expressions for the evolution of the AGN and galaxy populations  (derived from observations in a particular electromagnetic band),  and on simple scaling laws to relate the $\gamma$-ray luminosity with the properties of the host galaxies.

The paper is organized as follow. Section 2 describes the physical processes producing  $\gamma$-ray emission in AGN-driven and SN-driven shocks. A description of the SAM is given in Section 3. Section 4 describe how we model the $\gamma$-ray emission from AGN winds and star-forming galaxies. In Section 5 we derive the contribution to the EGB from AGN winds and star-forming galaxies ; Discussion and Conclusions follow in Sections 6 and 7.

\section{$\gamma$-ray emission from astrophysical shocks}\label{gamma_ray_emission}
The standard paradigm for the origin of the $\gamma$-ray emission in star-forming galaxies is non-thermal emission from relativistic  particles accelerated  in the shocks produced by SN explosions.  Similarly to the shocks surrounding SN remnants, the shocks produced by the interaction of  AGN winds with the surrounding ISM are expected to accelerate particles to relativistic energies \citep{Nims15,Wang16_gamma,Lamastra16}. 
In fact, outflows of ionized, neutral and molecular gas, extended from few milli-pc to kpc scales from the central supermassive black hole (SMBH) are now commonly observed  in local and high redshift AGN \citep[see][and references therein]{Fiore17}.  The most  powerful of these AGN winds are made by fast ($v\sim0.1-0.3 c$) highly ionized gas particles  that are likely  accretion disc  particles accelerated  by the AGN radiation field. 
The shock  pattern  resulting from the impact of a  AGN wind on the ISM gas  is similar to that of the stellar wind hitting the ISM around it (e.g.  \citealt{Weaver77,King15,King03,King03_2,King11,Lapi05,FGQ12,Zubovas12,Zubovas14}).
 The wind-ISM interaction  is expected to drive an outer forward shock into the ISM accelerating the swept-up material, and an inner reverse shock into the wind decelerating itself, separated by a contact discontinuity.  The cooling properties of the shocked wind gas determines whether the outflow is energy- or momentum-driven.  In the limit of efficient cooling of the shocked wind gas, most of the pre-shock kinetic energy is radiated away, and only its momentum flux is transferred to the ISM (momentum-driven). On contrast, if the  shocked wind gas does not cool, all the energy initially provided by the shock is retained within the systems, 
the shocked wind gas  expands adiabatically pushing the ISM gas away (energy-driven). 

Inelastic collisions between CR protons accelerated by AGN-driven  and SN-driven shocks with ambient protons may produce a significant   $\gamma$-ray emission. In fact, inelastic proton-proton collisions 
 produce neutral and charged pions. Neutral pions decay into two $\gamma$-rays: $\pi^{0} \rightarrow \gamma + \gamma$; while charged pions decay into secondary electrons and positrons and neutrinos: $\pi^{+} \rightarrow \mu^{+} + \nu_{\mu}$ and $\mu^{+} \rightarrow e^{+}+\nu_{e} +\overline{\nu}_{\mu}$; $\pi^{-} \rightarrow \mu^{-} + \overline{\nu}_{\mu}$ and $\mu^{-} \rightarrow e^{-}+\overline{\nu}_{e} +\nu_{\mu}$.
 CR electrons can also produce $\gamma$-ray emission either through interaction with ISM gas (bremsstrahlung) or interstellar radiation field (IC scattering). 

In our previous paper \citep{Lamastra16} we developed a physical model for  the  $\gamma$-ray emission from relativistic protons and electrons accelerated by astrophysical shocks. This model was used to predict the $\gamma$-ray spectrum produced by CR particles accelerated by the shocks observed in the molecular disk of the Seyfert galaxy  \textsc{NGC 1068}.
In this paper we derive the  gamma-ray emission from AGN winds and star-forming galaxies in a cosmological context  by including the physical model for  the  $\gamma$-ray emission into a semi-analytic model of hierarchical  galaxy formation. Our aim is to compare the model predictions with the measurement of the EGB  intensity  performed by the {\it Fermi}-LAT in the range between 100 MeV and 820 GeV \citep{Ackermann15}. In this energy range leptonic gamma-ray emission is expected to be lesser than the hadronic one , thus we limit the calculation of the $\gamma$-ray spectrum to the hadronic component.

Here we briefly recall the basic points of our model. We assume  that  protons are accelerated by diffusive shock acceleration (DSA)  to relativistic energies in the forward outflow shock. The resulting proton number density per unit volume can be expressed as a power-law with spectral index $p\simeq$2  and  an exponential high-energy cut-off  \citep{Bell78,Bell78b,Blandford78,Drury83}:
\begin{equation}\label{spectral_model}
N(E_p)=A_pE_p^{-p}\exp\left[-\left(\frac{E_p}{E_{max}}\right)\right].
\end{equation}
The normalization constant $A_p$ is determined by the total energy supplied to relativistic protons  at the shock, and $E_{max}$ is the maximum energy of accelerated protons.
The latter can be obtained by equating the proton acceleration time  $\tau_{acc}=E_pc/eBv_s^2$, where $e$ is the electron charge, $v_s$ is the shock velocity, and $B$ is the magnetic field strength,  to either  the time scale  of proton-proton collisions $ \tau_{pp} \approx 5 \times 10^7 yr/(n_H/cm^{-3}) $, where $n_H$ in the ISM number density, or the outflow time scale $\tau_{s}=R_{s}/v_{s}$. Thus:
\begin{equation}\label{Emax}
E_{max}=0.5v_{s,8}^2\tau_{age,3}B_{\mu G} \, TeV,
\end{equation} 
where $v_{s,8}$ is the shock velocity in units of 10$^8$ cm/s, $\tau_{age,3}$ is the age of the accelerator in units of 10$^3$ yr, $ B_{\mu G} $ is the magnetic field strength in units of $\mu$G \citep{Reynolds08}.\\
We constrain the normalization constant $A_p$ in eq. (\ref{spectral_model}) as:
\begin{equation}\label{Lkin_norm}
\int_{E_{min}}^{E_{max}} N(E) E dE = \eta_p  E_{kin},
\end{equation}
where  $E_{min}=m_pc^2$ is the  minimum energy of accelerated proton which is set to be proton rest mass, $E_{kin}$ is the kinetic energy of the shocked particles, and $\eta_p$ is the fraction of the  kinetic energy transferred to protons. For the latter we adopt $\eta_p\simeq$0.1 that is the valued  assumed in standard SN-driven shocks \citep {Keshet03,Thompson06,Tatischeff08,Lacki10}).

\subsection{Gamma-ray spectrum}\label{gamma_spectrum}

We compute the $\gamma$-ray spectrum produced by neutral pion decay using the $\delta$-functional approximation \citep{Aharonian00}:
\begin{equation}\label{L_gamma}
L_{\gamma}(E)=2VE^2\int_{Emin}^{\infty}\frac{q_{\pi}(E_{\pi})}{(E_{\pi}^2-m_{\pi}^2c^4)^{0.5}}dE_{\pi},
\end{equation}
where V is the volume of the outflow, $E_{min}=E+m_{\pi}^2c^4/4E$, and $E_{\pi}$ and $m_{\pi}$ are the energy and mass of the neutral pion. The emissivity of $\pi^0$  is given by:
\begin{equation}
q_{\pi}(E_{\pi})=\frac{cn_H}{k_{pp}} \sigma_{pp}(x)N(x),
\end{equation}
where $x=m_pc^2+E_{\pi}/k_{pp}$, $k_{pp}$=0.17 is the fraction of the accelerated proton energy that goes to neutral pions in each interactions, $\sigma_{pp}$ is the inelastic cross section of proton-proton collision, and $N(x)$ is the accelerated proton energy distribution.

\begin{figure*}[h!]
\begin{center}
\includegraphics[width=10 cm]{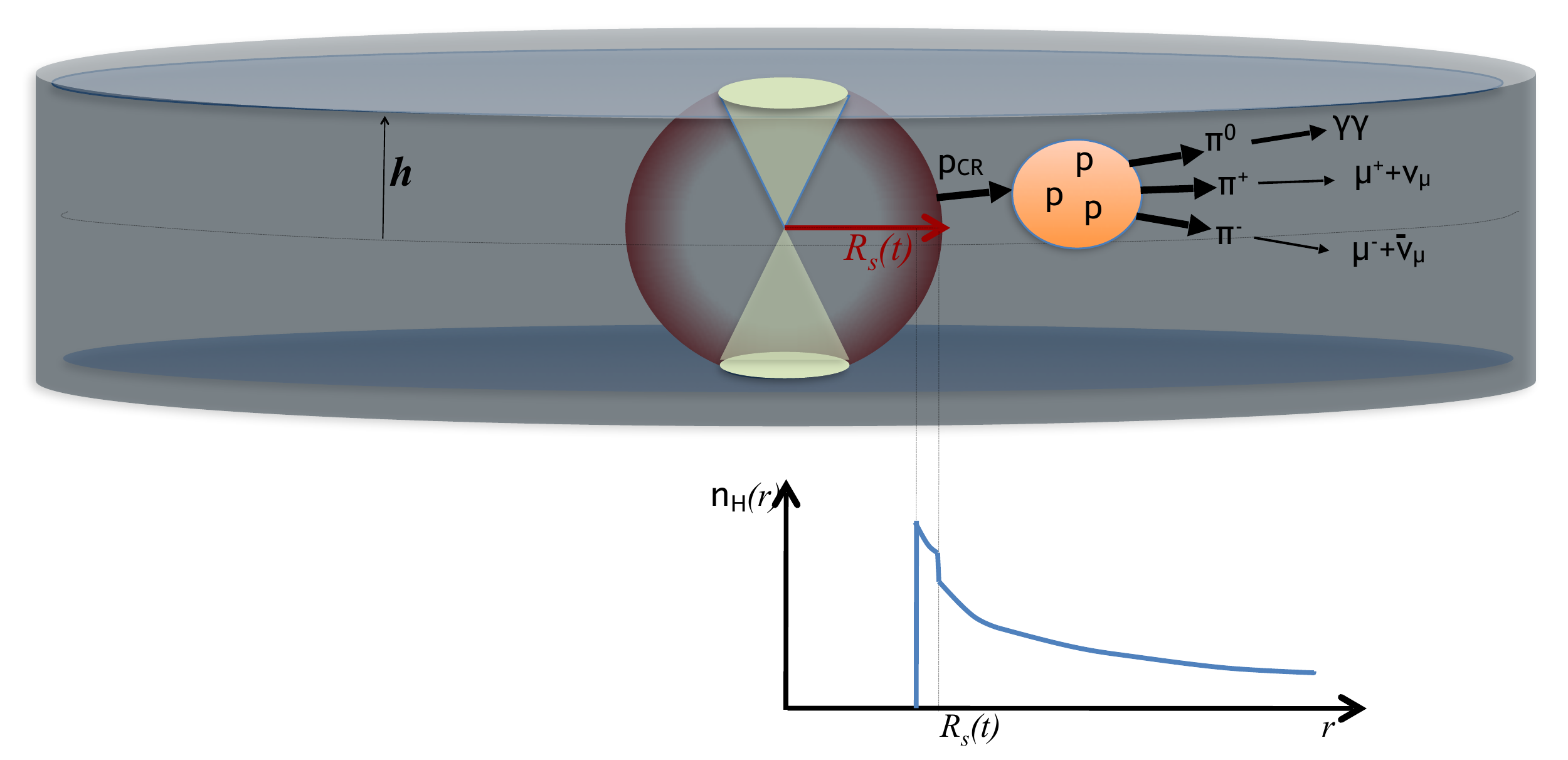}
\caption{Schematic representation of the interaction of CR protons accelerated in the AGN  blast wave with ISM protons. The shock radius $R_s(t)$ expands outwards  compressing the swept gas  into a thin shell, and leaving a cavity inside. The $\gamma$-ray emission from neutral pion decays occurs in  the regions outside the yellow cones where  the line of sights intercept  the galactic gas that has not been swept out by the blast wave.}
\label{disk}
\end{center} 
\end{figure*}

\section{The Semi-analytic model}\label{SAM}

In order to connect the above modelling of source emission to the statistical description of galaxy and AGN populations in a cosmological framework, 
we use the SAM described in  details in \cite{Menci14} (see also \citealt{Gatti15}). The SAM connects the cosmological evolution of  dark matter halos with the processes involving their baryonic content.  
An accurate Monte Carlo procedure is used to generate the merging trees of dark matter halos following the Press \& Schecter formalism   \citep{Bond91,Lacey93}, and  to  describe the gradual inclusion of sub-halos and their  dynamical friction processes and binary interactions (major and minor merging and fly-by events).

We assume a cold dark matter power spectrum of perturbations and we adopted  a Hubble constant $h$=0.7 in units of 100 km s$^{-1}$ Mpc$^{-1}$,  a dark energy density parameter $\Omega_{\Lambda}$=0.7, 
a matter density parameter  $\Omega_M$=0.3,  and a baryon density parameter $\Omega_b$=0.035.

The baryonic processes taking place in each  dark matter halos are computed following the standard recipes commonly adopted in SAMs. Starting from an initial amount $M_{gas}$=$M_{halo}\Omega_b/\Omega_M$ of gas at the virial temperature in each  dark matter halos,  we compute the mass of cold baryons which are able to radiatively cool. The cooled gas settles into a rotationally supported disk with mass $M_{c}$, disk circular velocity $v_d$, and disk radius $r_d$ (typically ranging from $1$ to $5 $ kpc) computed as in \cite{Mo98}. 

The cooled gas mass $M_{c}$ is converted into stars through two different channels: i) quiescient  star formation, gradually converting  the gas into stars  with a rate SFR= $M_{c}/\tau_*$ given by the Schmidt-Kennicutt law with $\tau_*$=1 Gyr; ii) starbursts following galaxy interactions (merging and fly-bys) occurring on time scales $ \sim 10^{7}-10^{8}$ yrs given by the duration of the interaction.

We assume that all stars with masses in the supernovae regime explode together, giving rise to a single bubble, and that a fraction of the total energy released by supernovae explosions is fed back onto the galactic gas. Thus, the effect of supernovae feedback is to  return part of the cooled gas into the hot phase. The mass $\Delta m_h $ returned from the cold gas content of the disk to the hot gas phase  is estimated, at each time-step, from canonical energy balance arguments \citep{Kauffmann96,Kauffmann98} as $\Delta m_h= E_{SN} \epsilon_{SN} \phi \Delta m_*/v_c^2 $, where $E_{SN}$=10$^{51}$ erg is the energy of ejecta of each supernova, $\epsilon_{SN}$=0.01-0.5 is the efficiency for the coupling of the emitted energy with the cold ISM, $\phi$=0.003-0.005 $M_{\odot}^{-1}$ is the number of supernovae per unit solar mass, depending on the assumed initial mass function (IMF), and $v_c$ is the circular velocity of the galactic halo. The model free parameter  $\epsilon_{SN}$=0.1 are chosen as to match the local $B-$band luminosity function and the Tully-Fisher relation adopting a Salpeter IMF. 
Although our simple modelling of supernovae feedback does not include a detailed treatment of the gas kinematics, including the dynamics of superbubbles \citep{Ferrara00}, it provides a good match to the observed correlations between the outflow velocity with the galactic circular velocity, and the SFR \citep{Calura09}.

The luminosity produced by the stellar population of the galaxies are computed by convolving the star formation histories of the galaxy progenitors with a synthetic spectral energy distribution (SED, \citealt{Bruzual03}). 
The dust extinction affecting the above luminosities is computed assuming the dust optical depth to be proportional to the metallicity $Z_{cold}$ of the cold phase (computed assuming a constant effective yield) and to the disk surface density, so that for the $V$ band $\tau_V \propto M_c Z_{cold}/\pi r_d^2$.  The proportionality constant is taken as to match the bright end of the local luminosity function. To compute the extinction at other wavelengths we applied a proper extinction curve \citep[see][]{Menci02,Menci05}.

The SAM includes the growth of supermassive black hole (SMBH)  from primordial seeds. The latter are assumed to be the end-product of  PopIII stars with a mass $M_{seed}$=100 $M_{\odot}$ \citep{Madau01}, and to be initially present in all galaxy progenitors. SMBH grow by merging with other black holes following the coalescence of the host galaxies and by accretion of cold galactic gas.  The latter gives rise to the AGN activity.  
The gas accretion  is triggered by  galaxy interactions. In particular, we assume the analytical description of the gas inflows induced by galaxy interactions derived by \cite{Cavaliere00} (see also  \citealt{Menci06,Menci08,Lamastra13b}), and  that in each galaxy interaction 1/4  of the destabilized gas feeds the  SMBH, while the
 the remaining fraction feeds the circumnuclear starburst \citep{Sanders96}.  These gas fractions are calibrated as to yield final SMBH masses matching the observed local correlations with the properties of the host galaxies.
We converted the mass accretion rate  $ \dot{M}_{BH} $ into AGN bolometric luminosity as:
\begin{equation}
L_{AGN}=\eta \dot{M}_{BH}c^2,
\end{equation}
where $\eta\simeq$0.1 is the  efficiency for the conversion of gravitational energy into radiation \citep{Yu02,Marconi04}. \\
Our SAM also includes a model of AGN feedback described in detail in Section  \ref{AGN_feedback}.\\

The SAM has been tested against the  statistical properties of the galaxy and AGN populations at low and high redshift and in different electromagnetic bands. 
In particular, the model provides galaxy luminosity function (LF) in the $K$-band  (a proxy for the stellar mass content) and in the UV band (a proxy for the instantaneous SFR)  that are in good agreement with the observed evolution of the  galaxy LF in the $K$-band up to $z\sim$3 and with the galaxy LF in the UV band up to $ z\sim $6 \citep[see][]{Menci14}.  The model is also able to reproduce the well known bimodal distribution of galaxies  in the color-magnitude diagram  \citep[see][]{Menci14}.\\ 
The model predicts AGN  LF  in the UV band  that are in good agreement  with the observational estimates at intermediate and high luminosities up to  $z\sim$6. At all redshift the model tends to slightly overestimate the data at faint luminosities \citep[see][]{Menci14}. 
The observational scaling relations between the galaxy and AGN physical properties  (such as stellar mass, SFR, SMBH mass, and $ \dot{M}_{BH} $) are also well described by the model  \citep{Menci05, Menci06,Lamastra10,Lamastra13a,Lamastra13b,Menci14,Gatti15}.\\

\subsection{The blast wave model for AGN feedback}\label{AGN_feedback}
Our SAM includes a physical model for AGN feedback which is related to the impulsive luminous AGN phase.
As discussed in Section \ref{gamma_ray_emission}, mildly relativistic winds ($v\sim $0.1-0.3$c$ ) are  injected by AGN into the surrounding ISM  \citep{Chartas02,Pounds03,Reeves03,Tombesi10,Tombesi15}.  
As these winds propagate into the ISM, they compress the gas into a blast wave terminated by a leading shock front, which moves outward with a lower but still supersonic velocity and sweeps out the surrounding medium.
The expansion of the blast wave into the ISM is described by  hydrodynamical equations. Taking into account the effect of dark matter gravity, upstream pressure, and initial density gradient, and assuming the Rankine-Hugoniot boundary condition at the shock, \cite{Lapi05} derived an analytic expression for the radius $R_{s}$ of the blast wave in the case of shock expansion in a gas with a power-law density profile $\rho\propto r^{-\omega}$, where the exponent $\omega$  is in the  range $2\leq \omega <2.5$  \citep[see also][]{Chevalier76,Chevalier82,Weaver77,Ostriker88,Franco91}. \\
In    \cite{Menci08} the expression for  the shock radius  is given  in terms of the galactic disk radius, disk velocity, and  Mach number ${\mathcal M}=v_s/c_s(R_s(t))$:
\begin{equation}\label{R_flow}
R_{s}(t)=v_d\,t_d\,\Big[{5\,\pi \omega^2 \over 24 \pi (\omega-1)}\Big]^{1/\omega}
{\mathcal M}^{2/\omega}\Big[{t\over t_d}\Big]^{2/\omega}.
\end{equation}
The Mach number ${\mathcal M}$  is  related to ratio between the energy $\Delta E$ injected by the AGN into the surrounding medium and the total thermal energy $E\propto M_{c}$ of the ISM:
\begin{equation}
{\mathcal M}^2=1+\Delta E/E.
\end{equation}
Thus the production of weak (${\mathcal M}\simeq$1) or strong shocks (${\mathcal M}\gg$1) depends on the value of  $\Delta E$  which is computed as:
\begin{equation}
\Delta E = \epsilon_{AGN}L_{AGN}\tau_{AGN}, 
\end{equation}
here $ \epsilon_{AGN} $ is the fraction  of the AGN bolometric luminosity transferred to the gas in the form of kinetic energy, and $\tau_{AGN}$  is the duration of the AGN phase.

The  blast wave model for AGN feedback was used in our previous papers to explain the distribution of hydrogen column densities in AGN as a function of luminosity and redshift, and to predict hydrogen phoionization rate as a function of redshift \citep{Menci08,Giallongo12}.

\section{Model set up}

In this Section we describe the model parameters that we will use in the computation of the $\gamma$-ray emission from AGN winds and star-forming galaxies.\\
In particular, we define  the parameters that describe the  $\gamma$-ray spectrum of individual AGN wind and star-forming galaxy, and the environment into which the shocks expand.
We limit the shock expansion into galactic disks, for which we assume  a constant scale height $h_d$=100 pc \citep{Narayan02,Kruit11}, and an isothermal gas density profile  $n_H=n_{H,0}/ r^{-2}$ (see figure \ref{disk}). The constant $n_{H,0}$ in the density profile  can be constrained by the total gas content in the disk $M_c$.

\subsection{Modelling $\gamma$-ray emission from AGN winds}\label{gamma_AGN}

To derive the $\gamma$-ray spectrum of an individual AGN wind, we need to determine the energy distribution of the particles accelerated in the shocks.
The latter is determined by the CR particle spectral index $p$, the particle maximum energy $E_{max}$, and by the total energy supplied to relativistic particles  at the shock.
As discussed in Section  \ref{gamma_ray_emission}, we assume that  DSA is the  mechanism which produces CR protons in  AGN-driven shocks (but see \citealt{Vazza15,Vazza16,vanWeeren16} for results showing that DSA  has difficulties in explaining the observed emissions of particles accelerated in some astrophysical shocks).
DSA could result in the production of a  power-law accelerated proton  population with a power-law index $p\simeq$2 \citep{Bell78,Bell78b,Blandford78,Drury83}. 
As the particle diffuse from the acceleration region, energy-dependent diffusion losses can soften the source spectrum leading to  larger values of the spectral index $p$. The accelerated particle energy distribution  extends to energies as high as is permitted by various loss processes. Protons are accelerated up to a maximum  energy  that  depends on the  shock velocity,  age of the accelerator, and  on the magnetic field strength in the shock region (eq. \ref{Emax}). The first two parameters are determined by the hydrodynamics of the shocks and by the density of the galactic disk.
For an isothermal gas disk,  the blast wave shock radius is given by  $ R_{s}=0.9\,{\mathcal M}\,v_d\,t $  (eq.  \ref{R_flow}),  which implies a constant outflow velocity $v_{s}=dR_s/dt=0.9{\mathcal M}v_d$. 
For the magnetic field  $B$ we assume values bracketed by a minimum value that is given by the volume average ISM magnetic field strength $B_{ISM}=6\times(\Sigma_{gas}/0.0025 g cm^{-2})^a \mu G$ where $a\simeq$0.4-1  and $\Sigma_{gas}=2 n_H m_H h$ is the disk gas surface density \citep{Robishaw08,Lacki10,Mcbride14}; and a maximum value that is derived by assuming that a fraction $\xi_B\simeq$0.1,  based on observation of SN remnants \citep{Chevalier98},  of the post shock thermal energy is carried by the magnetic field $B_{shock}=(8\pi\xi_B n_skT_s)^{0.5}$  where $n_s$ and $T_s$ are the post-shock density and temperature of the gas respectively. To derive the maximum value of the magnetic field  $B_{shock}$ we assume 
the temperature and density jumps  given by the  approximations  valid for very strong shocks as given in \cite{Lapi05}: $n_s\simeq 4 n_H$ and $T_s \simeq \mu m_p v_s^2/3k$, irrespective of  ${\mathcal M}$.

A constant outflow velocity corresponds  to  an energy-driven outflow in which the kinetic luminosity does not vary in time. In this case, the kinetic energy of accelerated protons  is simply given by the product of the outflow kinetic luminosity $L_{kin}$ and the residence time of the  particles  in the acceleration region $\tau_{res}$: 
\begin{equation}\label{Ekin}
E_{kin}=L_{kin}\tau_{res}.
\end{equation}
In the case of AGN-driven winds, we assume that  the outflow kinetic luminosity is a fraction $ \epsilon_{AGN} $  of the AGN bolometric luminosity: 
\begin{equation}\label{Lkin_AGN}
L_{kin}^{AGN}=\epsilon_{AGN}L_{AGN}.
\end{equation}
The ratio between the outflow kinetic power and AGN bolometric luminosity has been recently determined in a sample of 94 AGN  by \cite{Fiore17} to be in the range $\epsilon_{AGN}$=0.001-0.1.  
 The parameter $\epsilon_{AGN}$=0.01 is chosen within the observational range as to provide a good fit  of the bright-end of the  AGN luminosity function (see \citealt{Menci14}).

\begin{figure*}[h!]
\begin{center}
\includegraphics[width=8 cm]{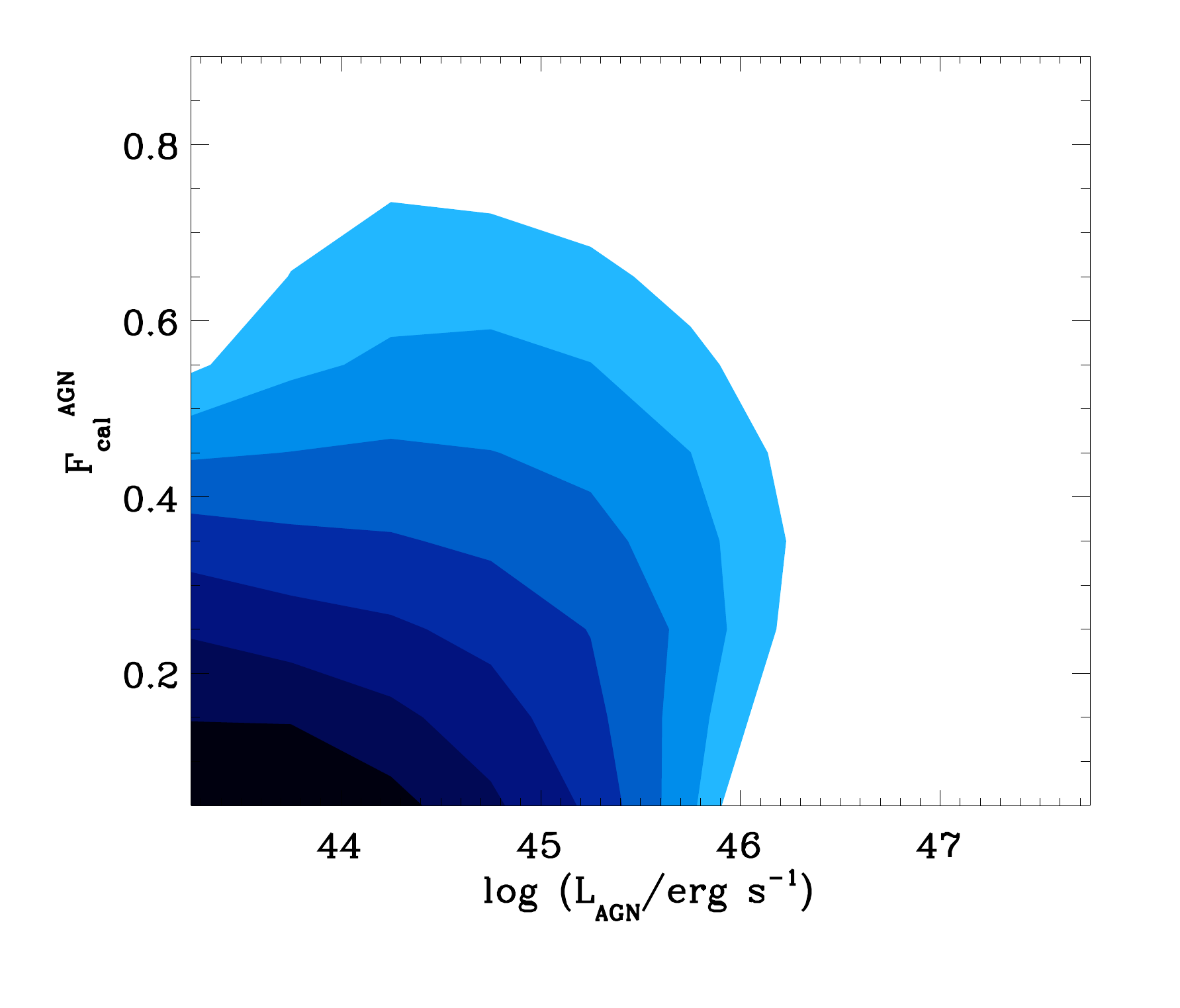}
\caption{Distribution of simulated AGN as a function of F$_{cal}^{AGN}$ and  AGN bolometric luminosity for our fiducial model. The contours correspond to equally spaced values of the density (per Mpc$^3$) of objects in a given F$_{cal}^{AGN}$-$L_{AGN}$ bin in logarithmic scale: from  $10^{-5}$ for the lightest filled region to $10^{-2}$ for the darkest.}
\label{Fcal}
\end{center} 
\end{figure*}

The collisions between  CR protons and ambient protons in galactic disks produce hadronic  $\gamma$-ray emission.
The blast wave model for AGN feedback allow us to self-consistently compute the fraction of accelerated protons that may interact with ambient protons. In fact  $\gamma$-ray emission from neutral pion decays occurs along the line of sights where the galactic gas has not been swept out by the blast wave produced by the AGN (see figure \ref{disk}).
This fraction is  the complementary of the escape fraction of ionizing photons that we derived in \cite{Giallongo12}.
The average of this fraction  over the duration $\tau_{AGN}$ of the AGN activity is given by:

 \begin{equation}\label{Fcal_geo}
 F_{cal}^{AGN} =\frac{\tau_s}{\tau_{AGN}}\left[1-\ln\left(\frac{\tau_s}{\tau_{AGN}}\right)\right],
\end{equation}
where   $\tau_s$ is the time at which the shock radius first encompasses the width of the galactic disk: $\tau_s=ht_d/0.9{\mathcal M}r_d$. For shocks with high Mach number, i.e. for large AGN injected energies $\Delta E>>E$,  $\tau_s << \tau_{AGN}$ yielding small fractions of interacting protons.  Figure \ref{Fcal} shows the distribution of simulated AGN as a function of $ F_{cal} ^{AGN}$ and  AGN bolometric luminosity for our fiducial model.

We assume that all the energy of the  protons that  interact with  the  protons in the galactic disk  is converted into  pion productions.  
This corresponds to assume that AGN winds act as proton calorimeters. The calorimetric limit has the maximum efficiency to convert AGN blast wave energy into $\gamma$-rays, and it  corresponds  to assume $\tau_{res}=\tau_{pp}$ in equation (\ref{Ekin}). 
The proton-proton collisional time scale $\tau_{pp}$ is  inversely proportional  to the density $n_H$ of target material, thus the resulting  hadronic $\gamma$-ray luminosity is independent on $n_H$, and it scales linearly  with  the outflow kinetic luminosity (see eq. \ref{L_gamma}).  Figure \ref{gamma_spectrum} shows the predicted $\gamma$-ray spectrum of a AGN with $L_{AGN}=$7$\times$10$^{44}$ erg/s  hosted  in a halo of mass 10$^{12} M_{\odot}$ at $z=$0.1. In our derivation of  the $\gamma$-ray spectrum  we neglect the $\gamma$-ray emission from IC and bremsstrahlung processes of the primary and  secondary leptonic populations, as it is expected to be lesser than the hadronic one at $E_{\gamma}\gtrsim$100 MeV \citep{Lacki14}.

\begin{figure}[h!]
\begin{center}
\includegraphics[width=8 cm]{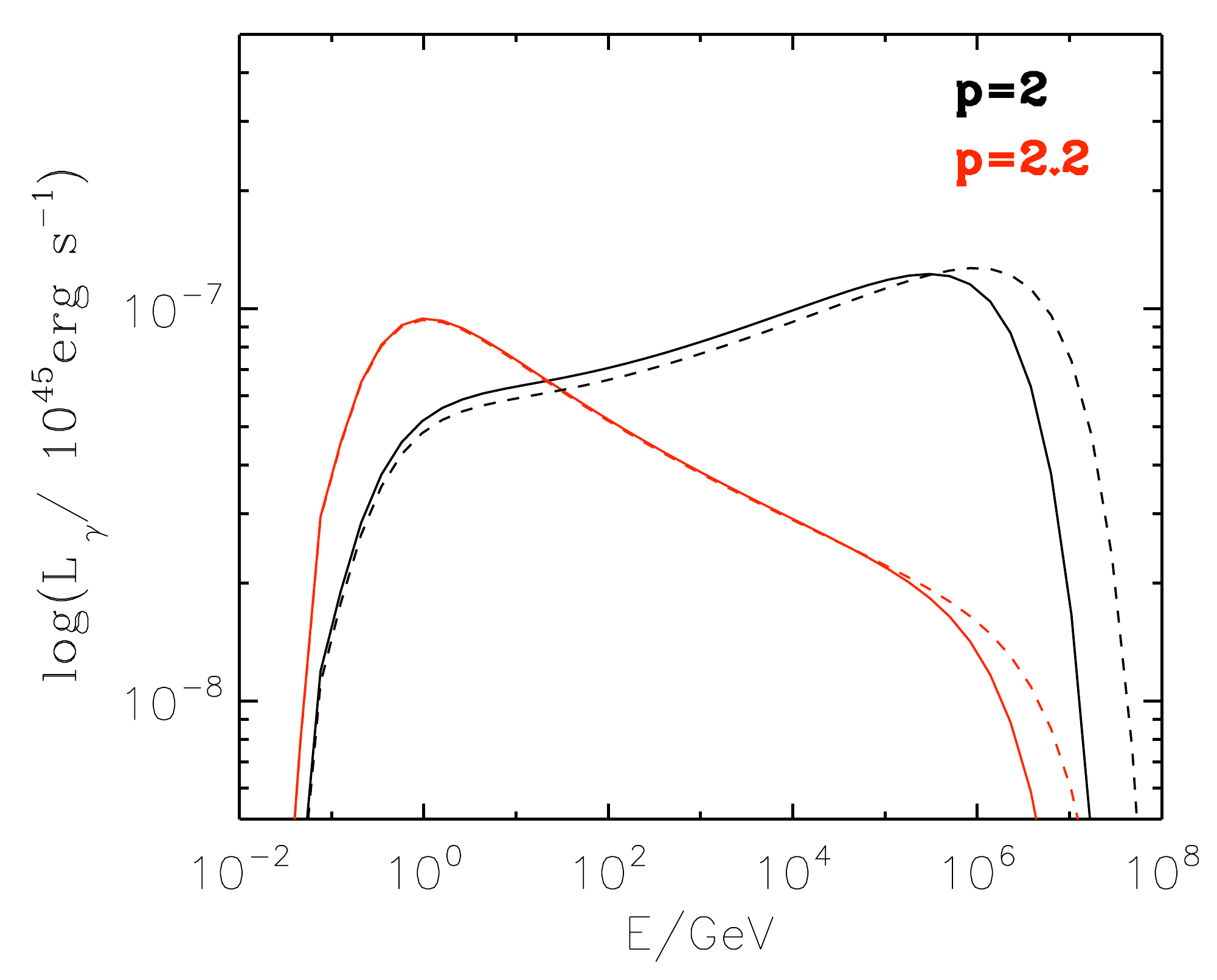}
\caption{$\gamma$-ray spectrum of an AGN with $L_{AGN}=$7$\times$10$^{44}$ erg/s  hosted  in a halo of mass 10$^{12} M_{\odot}$ at $z=$0.1. Energy spectrum parameter are set at $p=$2 (black lines), $p=$2.2 (red lines), $B=B_{ISM}$ (solid lines), and $B=B_{shock}$ (dashed lines).
 }
\label{gamma_spectrum}
\end{center} 
\end{figure}

\subsection{Modelling $\gamma$-ray emission from star-forming galaxies}\label{gamma_SF}

Beside the Milky Way, ten external star-forming galaxies have been firmly detected in gamma-rays with the Fermi-LAT \citep{Ackermann12,Ackermann17}. Among these, seven are starburst and active galaxies that are  more luminous at $\gamma$-ray energies compared to  quiescently star forming galaxies by a factor larger than 10.
The $\gamma$-ray spectra of starbursts look similar and can  be described by a single power-law with spectral index $ p \sim $2.2.
Because of the high density of star-forming regions in starburst galaxies, it is likely that they act as proton calorimeters \citep[][]{WangFields16}. Thus starburst galaxies  could  have  harder spectra than quiescent galaxies as in their acceleration region proton energy losses are  dominated by nearly energy-independent  proton-proton collisions.
On the contrary, in quiescent galaxies proton losses are thought to be dominated by energy-dependent diffusion  (as in the Milky Way), and their $\gamma$-ray luminosity is set by equilibrium between proton injection rate, diffusion processes, and energy losses. 
Both kind of star forming galaxies are included in our SAM. However, for the above reasons, modelling  the detailed escape and energy loss process of accelerated protons in quiescent galaxies is a very difficult task. Thus, in the following  we  will consider  only the calorimetric regime, which provide a good descriptions of  starburst galaxies, and we will  discuss the implications for the $\gamma$-ray emission from  the more numerous but less powerful quiescient galaxies.

To derive the energy distribution of  particles accelerated in starburst galaxies we assume a power-law index $p=2.2$ characteristic  of the  starbursts detected  in the $\gamma$-ray band \citep{Ackermann12}, we assume a constant shock  velocity equal to $v_{s}$=300 km/s \citep{Lacki10}, and a minimum and maximum value of the magnetic field in the shock region given by  $B_{ISM}$ and $B_{shock}$, respectively, as in the case of AGN-driven shocks.
The kinetic luminosity available in the form of accelerated protons for a starburst galaxy is: 
\begin{equation}\label{Lkin_SN}
L_{kin}^{SB}=\epsilon_{SN}\nu_{SN}E_{SN},
\end{equation}
where $\nu_{SN}$ is the supernovae rate, $E_{SN}\simeq 10^{51}$ erg is the typical energy from a supernovae explosion, and  $\epsilon_{SN}$=0.1  is the fraction  of the  SN energy transferred to the gas in the form of kinetic energy (see Section \ref{SAM}). Supernovae rate can be estimated from SFR and the IMF as  $\nu_{SN}=\phi\times SFR$  where $\phi\simeq$0.003  $M_{\odot}^{-1}$ for a Salpeter IMF.  We note that for starburst galaxies in the calorimetric regime the $\gamma$-ray luminosity scales linearly with the SFR.

 Starburst galaxies are selected from the SAM as model galaxies dominated by the SFR triggered by galaxy interactions. The latter at the same time trigger AGN activity.     
The starburst-AGN connection predicted by the SAM (see \citealt{Lamastra13b}) implies that the blast wave produced by the AGN  could sweep out the disk gas when the  starbursts are in actions. 
This hampers the derivation of  the fraction of protons accelerated in SN-driven shocks that interact with protons in the ISM. In fact, the latter depends  on the time delay between the trigger of the AGN and starburst activities, and on the starburst spatial distribution that are quantity that  can not be provided by the SAM. 
For this reason, in the following to derive the contribution from starburst galaxies to the EGB  we will adopt an empirical calorimetric fraction, which is derived from observations of star-forming galaxies in the  GeV band  \citep{Ackermann12}:
\begin{equation}
F_{cal}^{SB}=0.3\left(\frac{SFR}{M_{\odot}yr^{-1}}\right)^{0.16}\left(\frac{E_{SN}}{10^{51} erg}\right)^{-1}\left(\frac{\eta_p}{0.1}\right)^{-1}.\\
\end{equation}

\section{Results}\label{result}
In this section we derive the cumulative $\gamma$-ray emission from AGN winds and starburst galaxies predicted by the SAM.
The contribution from AGN winds (starburst galaxies)  to the EGB spectrum can be estimated as:
\begin{equation}\label{EGB_eq}
E^2\frac{dN}{dE}=\int_{0}^{z_{max}} \int_{L_{\gamma,min}}^{L_{\gamma,max}} \phi(L_{\gamma},z)\frac{I(E_{\gamma}^{'},L_{\gamma},z)}{4\pi D_{L}^{2}(z)}exp[-\tau_{\gamma \gamma}(E_{\gamma}^{'},z)]\frac{d^2V}{dz d\Omega}d L_{\gamma} dz,
\end{equation}
where $E_{\gamma}^{'}=E_{\gamma}(1+z)$ is the intrinsic photon energy, $\phi(L_{\gamma},z)$ is the comoving number density of AGN (starburst galaxies) per unit $\gamma$-ray luminosity as a function of redshift, $D_{L}(z)$ is the luminosity distance, $I(E_{\gamma},L_{\gamma},z)$  is the $\gamma$-ray spectrum of an individual AGN  (starburst galaxy) with integral $\gamma$-ray luminosity $ L_{\gamma} $  at redshift $z$,  the factor $d^2V/dzd\Omega$ represents the comoving volume element per unit redshift and unit solid angle, and $\tau_{\gamma \gamma}$ is the diffuse extragalactic background light (EBL) optical depth for photons with energy $E_{\gamma}$ at redshift of $z$. In fact, the emitted $\gamma$-rays, while travelling through the inter galactic medium, interact with the photons of EBL and get absorbed through $e+$/$e-$ pair production. The absorption probability increases with energy and distance of the $\gamma$-ray source.  
In our calculation we assume $z_{max}=$5 and adopt the EBL model of \cite{Stecker16}.

\begin{figure*}[h!]
\begin{center}
\includegraphics[width=14 cm]{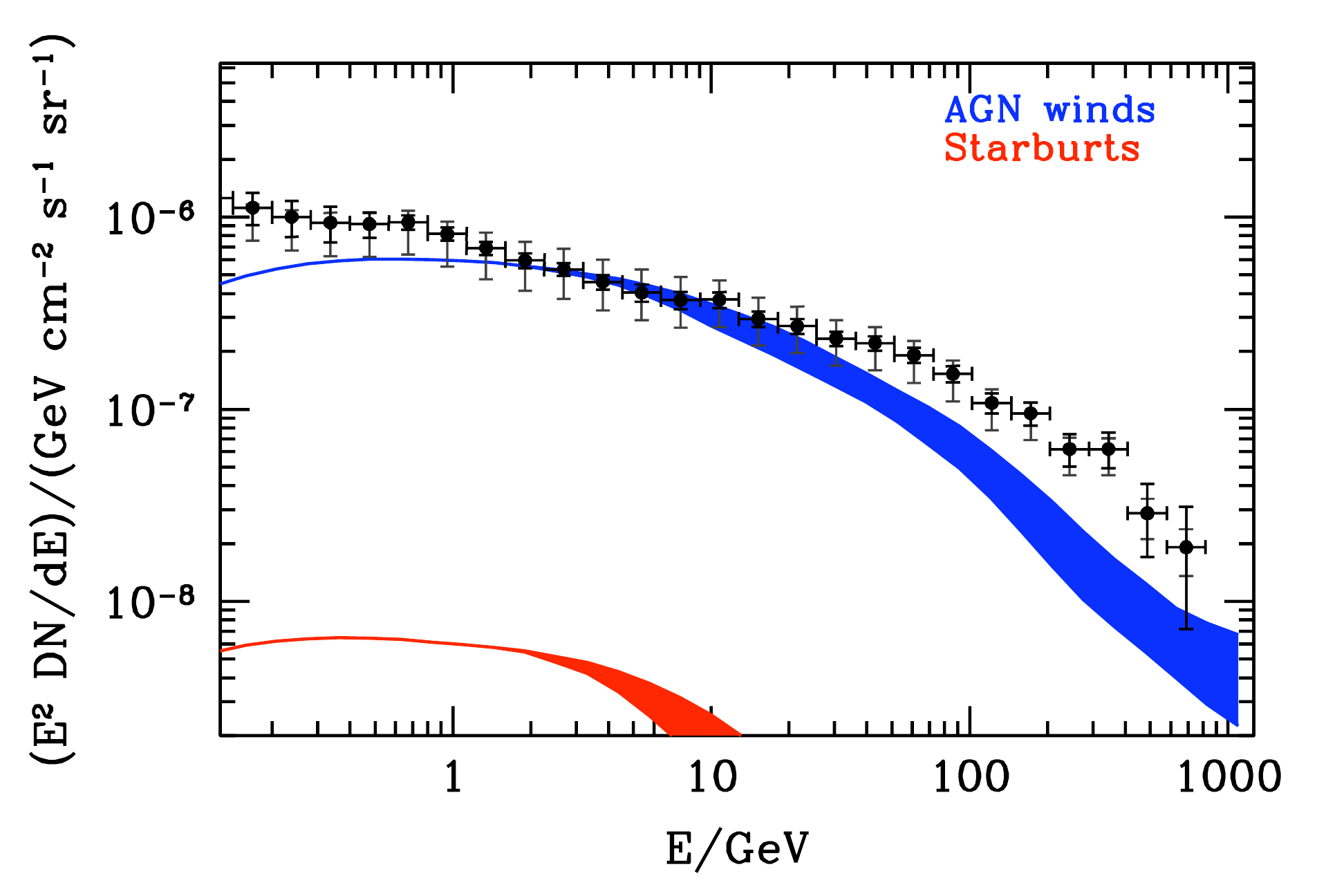}
\caption{The contribution to the  EGB from AGN winds (blue) and starburst galaxies (red) predicted by our SAM. Fiducial model parameters are set at: $h_d=$100 pc, $\eta_p$=0.1, $\epsilon_{AGN}=$0.01, and $\epsilon_{SN}=$0.1. Accelerated proton spectral index equal to  $p=$2.2 is assumed for both AGN winds and starburst galaxies. The blue and red shaded bands represent the uncertainty related to  the EBL model adopted \citep{Stecker16}. The data points are the  {\it Fermi}-LAT measurement of the EGB, with black and grey vertical error bars indicating EGB statistical and systematic uncertainties, respectively \citep {Ackermann15}.}
\label{back_gamma}
\end{center} 
\end{figure*}

Figure  \ref{back_gamma} compares  the cumulative $\gamma$-ray emission from AGN winds and starburst galaxies predicted by the SAM with  the EGB  spectrum measured by {\it Fermi}-LAT \citep{Ackermann15}. The latter  is well described by a power-law  with  spectral index of $\sim$2.3 and an exponentail cut-off.  Both the statistical and systematic uncertainties of the EGB measurement are  shown in figure \ref{back_gamma}. The  systematic uncertainty  ranges between a factor $ \sim $15\% and $ \sim $30\%  (depending on the energy range considered) and depends on the modelling of the Galactic diffuse emission.
The model predictions are shown for our fiducial model corresponding to  $h_d=100$ pc, $\eta_p=$0.1, $\epsilon_{AGN}$=0.01,  and $\epsilon_{SN}$=0.1. The accelerated proton spectral index is assumed to be  $p$=2.2 for both AGN winds and starburst galaxies. We find that our results do not depend on the exact value of the magnetic field. In fact, the magnetic field determines the cut-off energy of the EGB spectrum that is mainly affected by  EBL attenuation at the bright-end.
Figure \ref{back_gamma} shows that assuming a comparable efficiency for accelerating protons in AGN-driven and SN-driven shocks ($\eta_p=$0.1), hierarchical scenarios predict a  contribution to the EGB from AGN winds larger by about a factor of 100  than that  provided by starburst galaxies.
This implies  that the kinetic energy available to accelerate protons in  AGN winds  significantly exceed that provided by star formation driven by galaxy interactions.  
However, in hierarchical clustering scenarios starburst galaxies account for a small fraction of the cosmic star formation rate density  (SFRD), i.e. the mass converted into star per unit time and comoving volume. The starburst contribution to the SFRD was estimated to be $ \sim $5\% at $ z\simeq$0.1 and $ \sim $20\% at  $ z\simeq$5 \citep{Lamastra13a}.  Thus, the cosmic SFRD is dominated by quiescent galaxies at all redshift.  If we assume that, at all redshift, the $\gamma$-ray emission scales as the cosmic SFRD, we find that the contribution to the EGB from quiescent galaxies should be a factor of $\sim$20 greater than that 
provided by starburst galaxies. This implies that in hierarchical scenarios the  $\gamma$-ray emission from AGN winds dominate over that powered by  star-forming galaxies (quiescent plus starburst).
We note  that the above order-of-magnitude estimate for the quiescent contribution to the EGB should  be thought as an upper bound. In fact,  as discussed in Section  \ref{gamma_SF}, diffusive and advective losses cannot be neglected in quiescent galaxies, thus the hypothesis  of proton calorimeter adopted for  starburst galaxies, which has the maximum efficiency to convert supernova blast wave into $\gamma$-rays,   should not be valid in quiescent galaxies.

\begin{figure*}[h!]
\begin{center}
\includegraphics[width=10 cm]{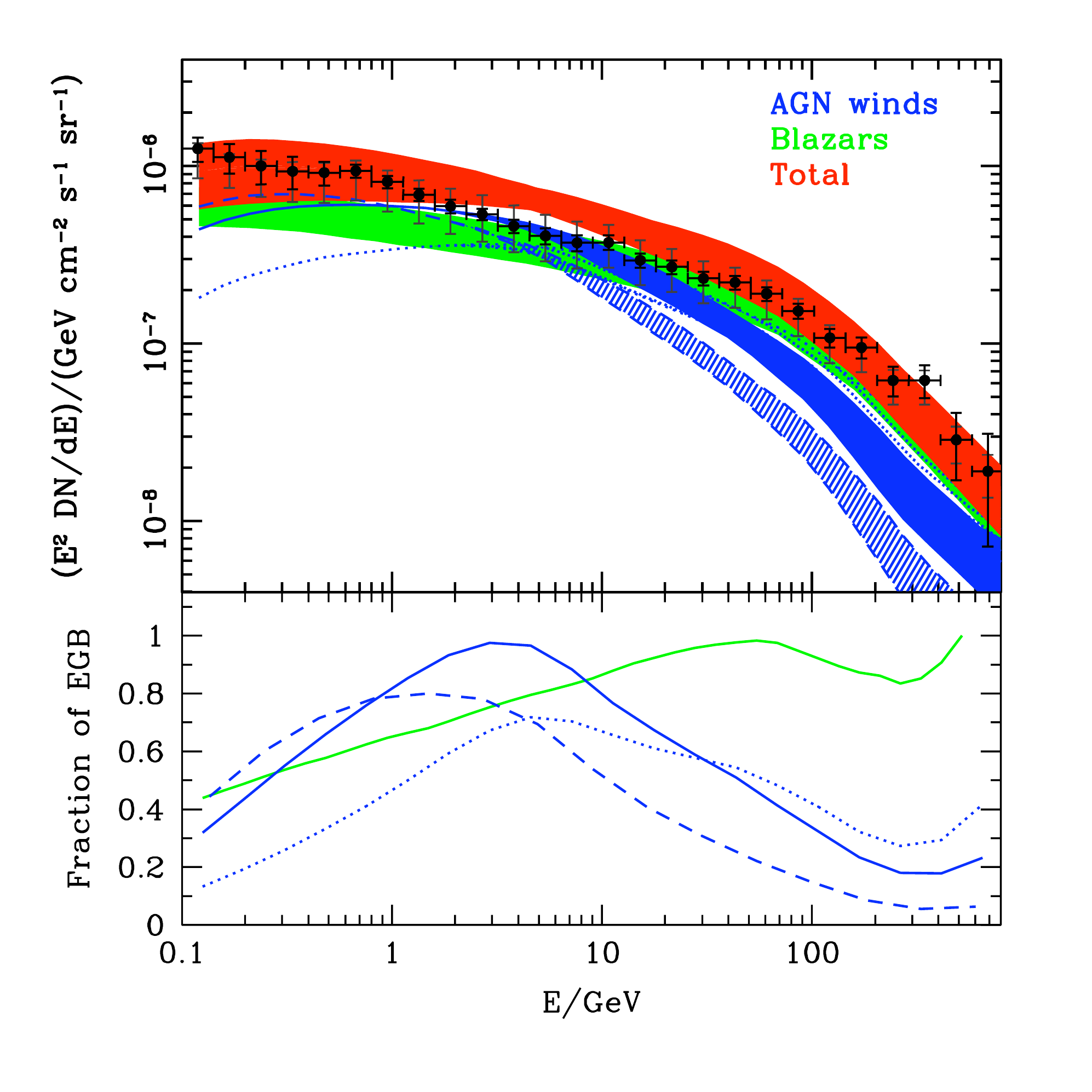}
\caption{Upper panel: the contribution to the EGB from AGN winds  predicted by our SAM is compared to the contribution provided  by the  blazar population \citep{Ajello15}. Model predictions are shown for our fiducial model and for different accelerated proton spectral index: $p=$2 (dotted blue lines), $p=$2.2 (solid blue lines), and $p=$2.4 (dashed blue lines).  The red shaded region represent the sum of the AGN wind and blazar contributions to the EGB.  The data points are as in figure \ref{back_gamma}. Lower panel: fraction of  EGB provided by AGN winds  and blazars.}
\label{missing_back_gamma}
\end{center} 
\end{figure*}

We find that AGN winds account for $ \sim $40\% of the observed EGB in the energy interval $E_{\gamma}=$0.1-1 GeV,  for $ \sim $90\% at $E_{\gamma}=$1-10 GeV, and for $ \sim $70\% at   $E_{\gamma}\gtrsim$10 GeV. Other classes of sources  are known to contribute to the EGB. Among these, the major contribution are from blazars.
The predicted cumulative emission of blazars  is shown in figure \ref{missing_back_gamma}  as green dashed band \citep{Ajello15}. The latter  encompasses systematic uncertainties on blazar  luminosity function models and  energy spectrum models.  We also show the integrated emission from AGN winds predicted by our SAM assuming different values of the accelerated proton spectral index. 
As shown in the bottom panel of figure \ref{missing_back_gamma},  at energies $E_{\gamma} \lesssim$10 GeV, and for spectral index $p>$2, the $\gamma-$ray emission from AGN winds dominates over that from blazars. The AGN wind contribution to the EGB is peaked at energies $E_{\gamma}\simeq$1-10 GeV, depending on the value of $p$, while the blazar contribution  reachs its maximum at larger energies $E_{\gamma}\simeq$60 GeV. 
This analysis shows that AGN winds and blazars can account for the amplitude and spectral shape of the EGB, leaving only little room for other contribution.

\section{Discussion}

\subsection{Robustness and comparison with previous works}
Here we discuss our assumptions and  compare our results with  previous estimates of the $\gamma-$ray emission from AGN winds and star-forming galaxies.

The results shown in the previous Section have been obtained assuming: i) energy-driven winds powered by the active nucleus in AGN host galaxies, ii) the calorimetric regime, and iii) shocks expansion limited to galactic disks.

As for the first point, the observations of winds  are very common in AGN host galaxies.
X-ray and UV emission and absorption line studies revealed outflows of highly ionized gas on  $ \sim $ 0.001-1 pc scale, with velocities $v=0.1-0.3 c$,   both at low and high redshift \citep[see e.g. ][]{Tombesi13,Tombesi15}. On larger scales (100-1000 pc), outflows of ionized, neutral atomic, and molecular gas, with velocities $v=10^2-10^3$ km/s   have been observed through deep optical/NIR spectroscopy and  interferometric observations in the (sub)millimetre band \citep[e.g.][]{Rupke11,Feruglio10,Feruglio15,Cicone14,Harrison16,Shen16,Zakamska16}.
A collection of  AGN winds detected at different scales and ionization states is given in \cite{Fiore17}. These observations indicate that the majority of the large-scale outflows are driven by the nuclear activity, and may be identified with the  energy-driven phase. In particular for two sources, namely MrK231 \citep{Feruglio15} and IRAS F11119 \citep{Tombesi15}, both X-ray winds and molecular winds have been detected. The comparison between the momentum rate of the X-ray and molecular winds indicates  that these winds are  energy-driven.
These findings seem to support our assumptions about the frequency and nature of AGN winds, however,  these observations are sparse, and mostly limited to AGN selected to have high chances for being in an outflowing phase. In order to gain more inside into this topics the measurements of the frequency and parameters of AGN winds  in unbiased AGN samples over a large range of redshift are  necessary.

As for the calorimetric regime, our model assumes that advective and diffusive escape of  accelerated protons in galactic disks are negligible. This condition is satisfied if the proton-proton collisional  time scale $\tau_{pp}$ is less than the advective and diffusion time scales. 
The wind advection time is $\tau_{wind}\simeq h_d/v_s$.
Assuming the velocities measured in galactic scale AGN winds, and for $h_d=100$ pc, we expect that hadronic losses dominate advection losses when $n_{H}\gtrsim$ 50-500 cm$^{-3}$, which are  values observed in circumnuclear disks of active and starburst galaxies \citep[e.g.][]{Garcia14,Yoast14} . 
The particle diffusion processes in the environments of  active and starburst galaxies are poorly constrained. For this reason we neglect diffusive losses, and this  could constitute a source of uncertainty in our computation.

As for the environment into which the shocks expand, we limit the expansion to galactic disks, however it is also possible that shocks can propagate in gaseous  haloes of galaxies.  The lower densities of galactic haloes with respect to galactic disks imply a major role of escape (advective and diffusive) with respect to hadronic interactions, and therefore a low $\gamma-$ray production rate. 
As a check, we can compare our predictions with other recent predictions of the EGB from AGN winds that assume shock expansion in both disk and halo component of galaxies.
\cite{Wang16_gamma}  derived the  cumulative $\gamma-$ray emission from AGN winds using a  hydrodynamical model for  AGN wind interaction with the ambient medium  \citep{Wang15},  and the empirical AGN bolometric luminosity function of \cite{Hopkins07}.  In their computation, the $\gamma-$ray emission  is assumed to be produced by the same hadronic processes considered  in this work and  described in Section \ref{gamma_ray_emission}.

\begin{figure*}[h!]
\begin{center}
\includegraphics[width=15 cm]{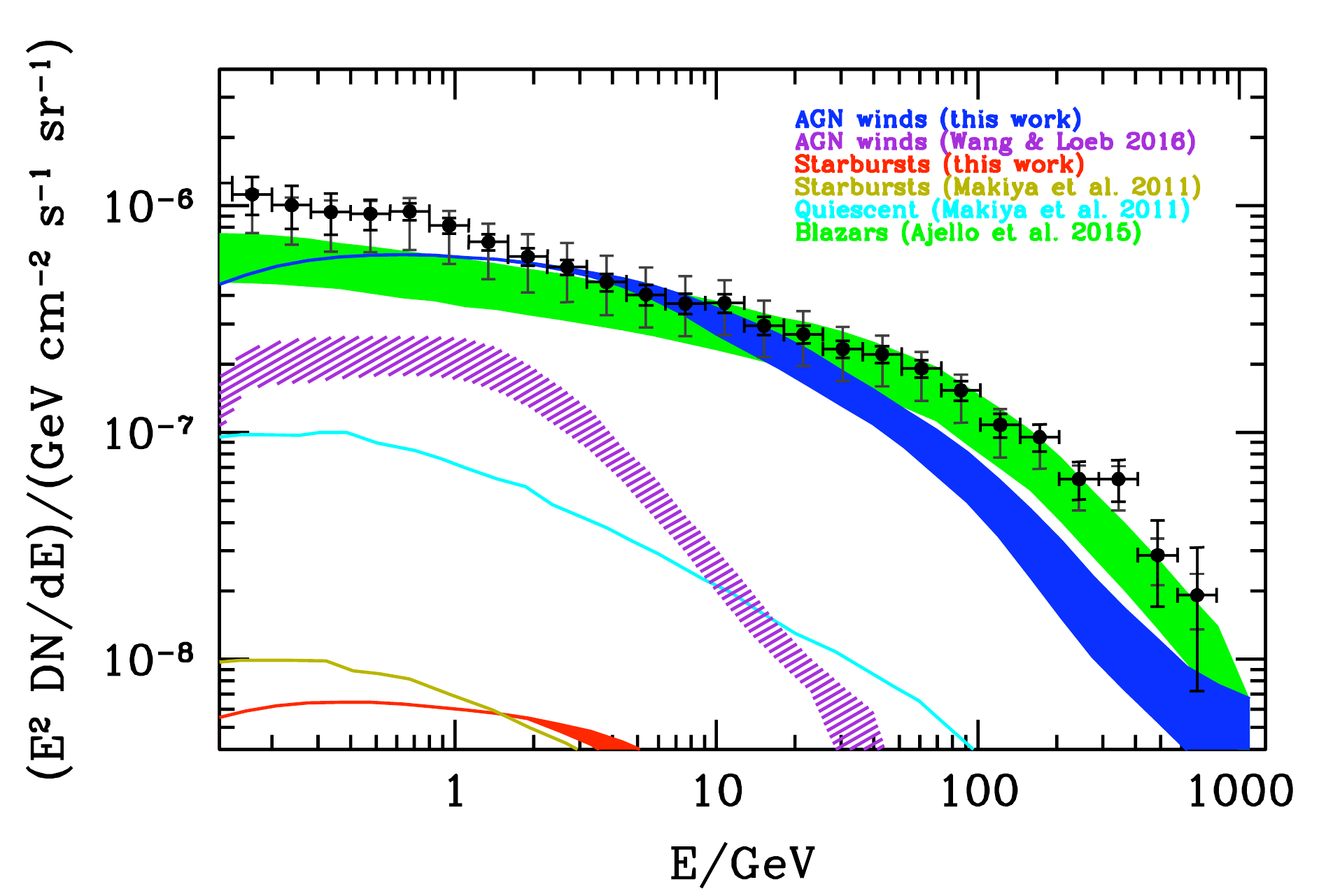}
\caption{EGB spectrum.  The blue and red  shaded regions correspond to AGN winds and starburst galaxies predicted by our SAM, respectively. The purple shaded region corresponds to the contribution to the EGB  from AGN winds  estimated by \cite{Wang16_gamma}. The cyan and gold lines indicate the estimate for quiescent and starburst galaxies predicted by the \cite{Makiya11} SAM, respectively. The green shaded region shows the contribution to the EGB  from blazars as estimated by \cite{Ajello15}. The data points are as in figure \ref{back_gamma}. }
\label{back_gamma_cont}
\end{center} 
\end{figure*}

 Figure \ref{back_gamma_cont} compares the estimates for the cumulative $\gamma-$ray emission from AGN winds derived by \cite{Wang16_gamma} with that predicted by our SAM assuming the same fraction of the AGN bolometric luminosity that powers the winds, and the same energy spectrum. 
 Figure \ref{back_gamma_cont} shows that the  normalization  of the EGB spectrum predicted by the SAM is a factor of $ \sim $3 larger  than that derived by \cite{Wang16_gamma}. The SAM also predicts a different shape of the EGB spetrum at energies $E_{\gamma} \gtrsim$1 GeV.
As for  the normalization, we have checked whether  the  mismatch seen in figure \ref{back_gamma_cont} stems from the different AGN luminosity functions adopted. 
 \cite{Wang16_gamma} used the  AGN bolometric luminosity function of \cite{Hopkins07} which is derived by combining  measurements of AGN number density in  IR, optical, and X-ray bands, in the redshift interval $z=$0-6. 
Deriving the bolometric luminosity function  from the observed luminosity functions in different electromagnetic bands is not a trivial procedure as it requires assumptions on the bolometric corrections, and corrections for obscured sources.  \cite{Hopkins07} use a luminosity-dependent bolometric correction, and they correct their data for extinction and the fraction of Compton-thick AGN missed in IR, optical, and X-ray data.   They also give analytical approximations of the empirical bolometric luminosity function. In their calculation, \cite{Wang16_gamma} use  the pure luminosity evolution (PLE) model given by  \cite{Hopkins07} to describe the evoluzion of the AGN population. As discussed by \cite{Hopkins07}, although the PLE   model provides a reasonable lowest order of magnitude approximation to the data, it underpredicts the abundance of low-luminosity AGN at $z\lesssim$0.5. Such an underprediction of faint objects  explains the normalization difference seen in figure \ref{back_gamma_cont}.

As for the shape of the EGB spectrum,  we find that  while at energies $E_{\gamma}\lesssim$1 GeV the two spectral shapes are consistent, at higher energies the SAM predicts a decline in the EGB spectrum
that begins at  larger energies and  is shallower than that predicted by \cite{Wang16_gamma}. The high-energy part of the predicted EGB spectrum is  affected by  $\gamma-$ray absorption in the intergalactic medium due to the EBL. There are large uncertainties regarding the  EBL estimates and thus also on the $\gamma-$ray optical depth used in equation (\ref{EGB_eq}) to calculate the EGB spectrum. We used  $\tau_{\gamma \gamma}$ given by \cite{Stecker16}, while \cite{Wang16_gamma} used the $\gamma-$ray optical depth estimated by the same authors in a previous paper \citep{Stecker07}. The latter results in a larger absorption of  $\gamma-$rays at energies $E_{\gamma}\gtrsim$10 GeV, and this could in part explain the discrepancy in the shape of the EGB  spectra seen in figure \ref{back_gamma_cont}. In fact, we have verified that assuming  $\tau_{\gamma \gamma}$ given by \cite{Stecker07},  the SAM predicts a steeper decline of the cumulative $\gamma-$ray emission from AGN winds, however the predicted EGB spectrum still remains shallower that that predicted by \cite{Wang16_gamma}. \\
It is also worth to note that the solutions of the hydrodinamical equations describing the outflow motion in an isothermal gas density profile give a constant outflow velocity (i.e. an energy-driven outflow) in the \cite{Lapi05} model, and an outflow velocity that decreases with increasing outflow radius (i.e. a momentum-driven outflow) in the \cite{Wang16_gamma} model. The shock velocity determines the maximum energy of accelerated protons, which in turn shapes the high-energy part of the EGB spectrum. The maximum energy that can attain accelerated protons can be expressed in terms of the shock velocity, size of the accelerator, and magnetic field strength as $E_{max}\propto v_s R_s B$ (see eq. \ref{Emax}). Since $B_{shock} \propto R_s^{-1}$ for an isothermal gas density profile, the maximum energy scale as  $E_{max}\propto v_s$.   Thus in an energy-driven outflow the  proton maximum energy remains constant during the outflow expansion, while in a momentum-driven outflow $E_{max}$ decreases with time.\\

As for the contribution to the EGB from star-forming galaxies,  \cite{Makiya11} derived the contribution  from both starburst and quiescent galaxies using a SAM of hierarchical galaxy formation.  As in our model, the star formation in starburst galaxies is triggered by galaxy interactions, while in quiescent galaxies the star formation is  determined by the cold gas reservoir and the galaxy dynamical time-scale. In the \cite{Makiya11} model the emission of quiescent and starburst galaxies are modelled based on templates which are tuned to reproduce the  $\gamma-$ray spectra  of Milky Way and  M82, respectively. For quiescent galaxies they assumed the so-called escape regime. In this regime the energy losses of accelerated protons are dominated by their escape from the diffuse region of the galaxy, and the  $\gamma-$ray luminosity depends on SFR and mass of ISM gas, i.e. $L_{\gamma} \propto$ SFR $\times M_{c} $. 
For the starburst galaxies they assumed  the calorimetric regime. In this case the $\gamma-$ray luminosity is no longer dependent to the gas mass, i.e. $L_{\gamma} \propto SFR $. 
The authors determined the relation between $L_{\gamma}$ and  SFR $\times M_{c} $, and $L_{\gamma}$ and  SFR, by fitting the results of 4 star-forming galaxies detected by 
{\it Fermi}-LAT. 
The contributions to the EGB from quiescent and starburst galaxies derived by \cite{Makiya11} are shown in figure \ref{back_gamma_cont}.  They found that the contribution to the EGB from quiescent galaxies is a factor of $ \sim $10  larger than that provided by starburst galaxies. The latter agrees reasonably well with our estimate. This  supports the scenario discussed in Section \ref{result} where  the   $\gamma$-ray emission from star-forming galaxies (quiescent plus starburst) predicted by  hierarchical galaxy formation models  is lesser than that provided by  AGN winds.

\subsection{Multi-messenger implications}
\begin{figure}[h!]
\begin{center}
\includegraphics[width=8 cm]{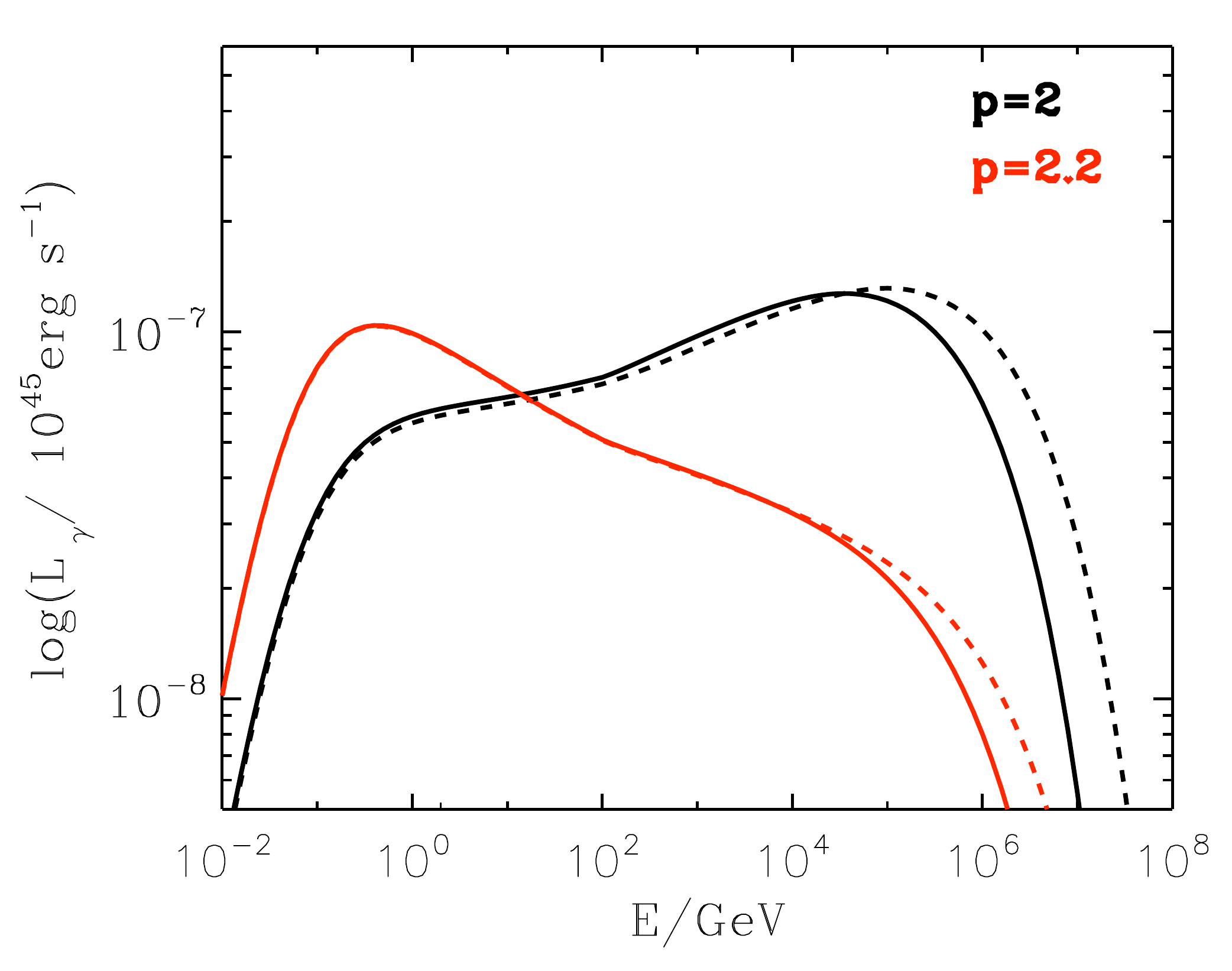}
\caption{Neutrino spectrum of an AGN  with $L_{AGN}=$7$\times$10$^{44}$ erg/s  hosted  in a halo of mass 10$^{12} M_{\odot}$ at $z=$0.1. Model parameter as in figure \ref{gamma_spectrum}.}
\label{neutrino_spec}
\end{center} 
\end{figure}

\begin{figure*}[h!]
\begin{center}
\includegraphics[width=15 cm]{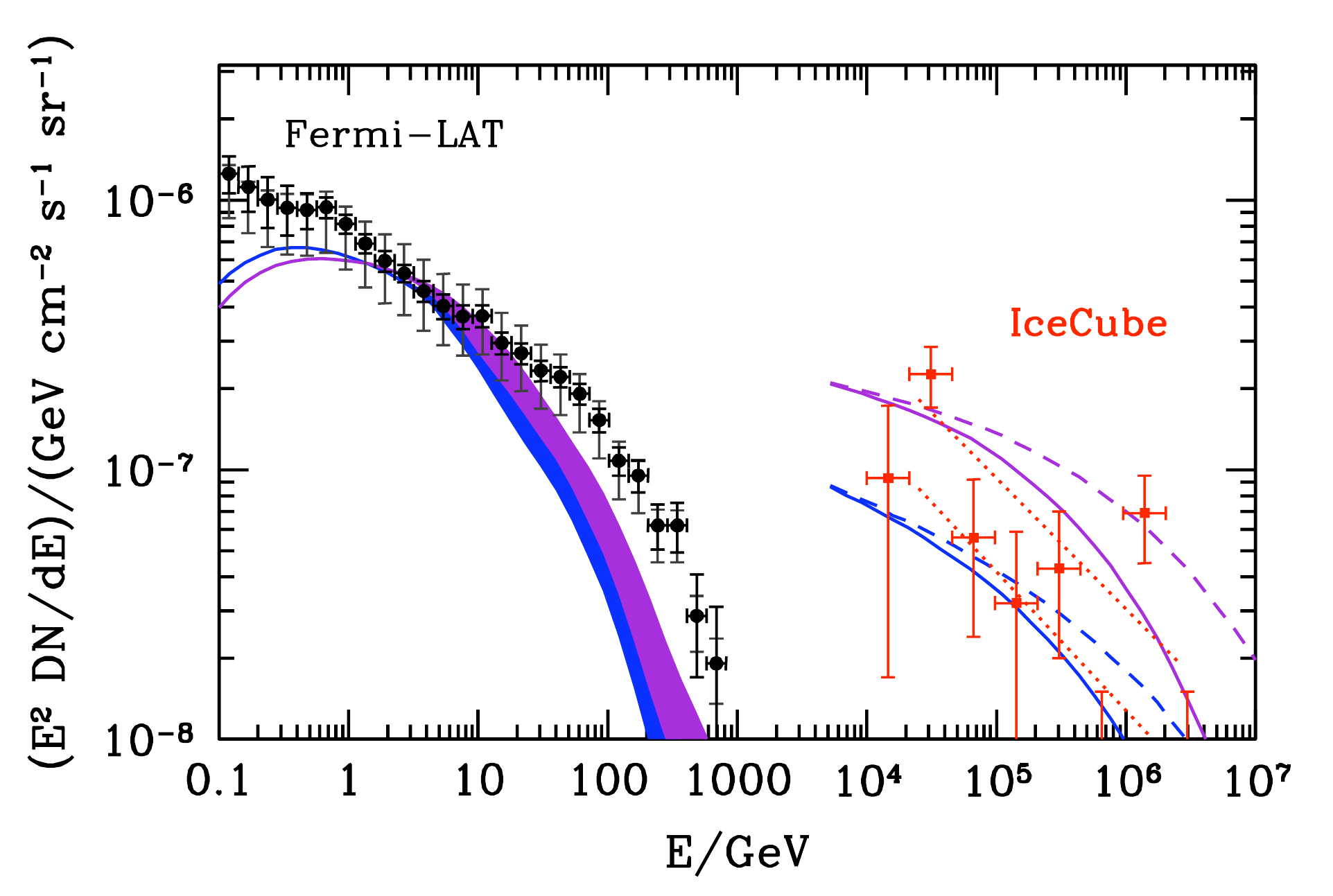}
\caption{Cumulative $\gamma$-ray (left) and neutrino (right) background from AGN winds predicted by our SAM. Model predictions are shown for our fiducial model and for different AGN wind spectral energy parameters: $p=$2.2 (purple lines), $p=$2.3 (blue lines),  $B=B_{ISM}$ (solid lines), and $B=B_{shock}$ dashed lines. The circles represent  {\it Fermi}-LAT data \citep{Ackermann15}. The  squares show the differential model of  IceCube neutrino data, while purple dotted lines represent the power-law models \citep{Aartsen15}. }
\label{nu_back}
\end{center} 
\end{figure*}

In  Section \ref{result} we have shown that a sizeable fraction of the EGB can be accounted by hadronic $\gamma$-ray emission from AGN winds.
The hadronic origin of the $\gamma$-ray flux will make the production of neutrinos unavoidable, creating a diffuse neutrino  background. Indeed, in  proton-proton interactions  $\sim$2/3 of the pions produced are  charged pions that decay into   muons and neutrinos followed by electrons and positrons and more neutrinos: $\pi^{+} \rightarrow \mu^{+} + \nu_{\mu}$ and $\mu^{+} \rightarrow e^{+}+\nu_{e} +\overline{\nu}_{\mu}$; $\pi^{-} \rightarrow \mu^{-} + \overline{\nu}_{\mu}$ and $\mu^{-} \rightarrow e^{-}+\overline{\nu}_{e} +\nu_{\mu}$. Neutrinos can also be created  in interactions of CR protons with the ambient radiation field. Here we focus on proton-proton interactions as they are the dominant process for AGN winds \citep{Wang16_neutrini,Lamastra16}.\\
To calculate the energy spectra of neutrinos produced in AGN winds  we use the parametrizations derived by  \cite{Kelner06} (see figure \ref{neutrino_spec}). Then,  the cumulative neutrino flux from AGN winds can be obtained by summing the neutrino emission over the entire AGN population, as described by eq (\ref{EGB_eq}). 
The absorption due to the EBL is ignored in the calculation of the neutrino background, and this introduces a dependence of the predicted spectrum on the values of the magnetic field in the shock region.\\
Figure \ref{nu_back} compares  the cumulative neutrino background from AGN winds with the most recent IceCube data. The latter are fitted by two different model: a single power-law model, and a differential model with nine free parameters  \citep{Aartsen15}. 
We find that the estimated neutrino intensity is comparable, within the astrophysical uncertainties, to the IceCube measurement  for spectral index $p\sim$ 2.2-2.3.
As shown in figure \ref{missing_back_gamma}, these spectral indexes imply  the largest contribution to the EGB from AGN winds at energies $E_{\gamma} \simeq$1-10 GeV, which, when added to the contribution from the balazar population, lead to a slightly overestimate of   {\it Fermi}-LAT data. Of course, any plausible model for the IceCube neutrino background should not  overpopulate the gamma bounds. This tension is pointing to a class of sources that are opaque in the $\gamma$-rays \citep{Chakraborty16,Murase16}. As discussed in \cite{Chakraborty16}, the large photon number density present in the environment of starburst galaxies  imply that  $\gamma$-rays could interact inside the galaxy before escape. The AGN-starburst connection predicted by our SAM imply that the internal absorption of $\gamma$-rays  could be present also in AGN host galaxies. Taking into account the internal absorption of $\gamma$-rays,  the {\it Fermi}-LAT and IceCube data could be  reproduced  simultaneously by our SAM \citep[see also][]{Wang16_neutrini}.

The analysis of positional coincidence of IceCube neutrino events with known astrophysical sources is a difficult task owing to the poor angular resolution of the detector. For this reason, there are as yet no  confirmed identifications for  astrophysical sources of IceCube neutrino events  \citep[e.g.][]{Aartsen14, Adrian16,IceColl16}. Recently,
\cite{Padovani16} (see also \citealt{Resconi17}) have argued for a statistical significant correlation between IceCube neutrino events and high energy peaked BL Lacertae (HBL) objects in the second catalogue of hard  {\it Fermi}-LAT sources \citep{Ackermann16}. Although HBL are promising neutrino-emitters candidates \citep{Lucarelli17}, they   can account for   $\sim$10\%-20\% of the IceCube signal.

In the next future the better angular resolution of KM3NeT  ($ \sim $0.2$^{\circ}$ for neutrinos with energy $E  \gtrsim $10 TeV, track-like events, \citealt{KM3Net_perf16}) will allow us to constrain  effectively the  position of the possible counterparts of neutrino events, thus providing a possible direct test of neutrino background models.\\

\section{Conclusions}

We have incorporated the description of the hadronic $\gamma$-ray emission from relativistic  protons accelerated in AGN-driven and SN-driven shocks into a state-of-the-art SAM  of  hierarchical galaxy formation. Our SAM has already proven to match  the  statistical properties of the galaxy and AGN populations at low and high redshift and in different electromagnetic bands.  We have compared the predictions for the cumulative $\gamma$-ray emission from AGN winds and star-forming galaxies with the latest measurement of the EGB  performed by the {\it Fermi}-LAT in the range between 100 Mev and 820 GeV \citep{Ackermann15}.
The main result of this paper are:
\begin{itemize}
\item in hierarchical clustering scenarios, connecting the physics of AGN and starburst galaxies to the merging histories of the host galaxies, assuming a comparable efficiency for accelerating protons in AGN-driven and SN-driven shocks ($\eta_p=$0.1),  the contribution to the EGB from AGN winds dominate over that from starburst galaxies. If we consider also the contribution of the less powerful but more numerous quiescent galaxies, the contribution to the EGB from all star-forming galaxies is a factor $ \sim $3-5 lower than that provided by AGN winds.

\item The cumulative $\gamma-$ray emission from AGN winds and blazars can account for the amplitude and spectral shape of the EGB, assuming the standard acceleration theory, and AGN wind parameters  that agree  with observations.  
At energies lower and greater than $E_{\gamma} \simeq$10 GeV the EGB is dominated by AGN winds and blazars, respectively.
The transition between these two regimes could, in principle, gives rise to breaks and features in the EGB energy spectrum. 

\item The neutrino background resulting from charged pion decays following hadronic interactions can reproduce the IceCube data assuming accelerated proton spectral  index $p\sim$ 2.2-2.3. 
The {\it Fermi}-LAT   data could be  reproduced simultaneously, taking into account internal absorption of $\gamma$-rays.

\end{itemize}


\bibliographystyle{aa}
\bibliography{biblio.bib}

\end{document}